\documentclass[aps,prl,twocolumn,superscriptaddress,floatfix,nobalancelastpage]{revtex4-2}
\usepackage{amsmath,graphicx}
\usepackage[colorlinks=true,linkcolor=blue,citecolor=blue,urlcolor=blue,pdftitle={Higher-order interactions renormalize the thermodynamic signature of neural population codes},pdfauthor={Narendra N. Hegade}]{hyperref}
\graphicspath{{./}}
\makeatletter\AtBeginDocument{\def\bibfont{\footnotesize\@clubpenalty\clubpenalty}}\makeatother
\begin{document}

\title{Higher-order interactions renormalize the thermodynamic\\ signature of neural population codes}

\author{Narendra N.\ Hegade}
\email{narendrahegade5@gmail.com}
\affiliation{IDAL, Electronic Engineering Department, ETSE-UV, University of
Valencia, Avgda.\ Universitat s/n, 46100 Burjassot, Valencia, Spain}
\affiliation{Kipu Quantum, Greifswalderstrasse 212, 10405 Berlin, Germany}

\begin{abstract}
Maximum-entropy models of neural recordings can be treated as Boltzmann distributions at a fictitious temperature $T$, with the fitted model at $T=1$. Their specific-heat peak grows with population size and approaches that temperature, which has been read as a sign of criticality. The case rests on pairwise models, yet retinal codes require higher-order interactions. The thermodynamics of a model with such interactions fitted cell by cell has not been computed. Here I compute it exactly for pairwise and third-order models of the same populations of up to $20$ retinal ganglion cells, in salamander and marmoset. The triplet terms, required by the held-out likelihood, shift the peak by more than the finite-sample bias, with opposite sign in the two retinas. The peak is lower and grows more slowly in salamander, higher and grows faster in marmoset, and lies closer to $T=1$ in both. An exact decomposition traces most of the shift to the energy variance across population spike counts. A constraint on the spike-count distribution alone reproduces all of it in salamander and over half in marmoset. A second recording of each species, rebinning, and a common-input surrogate show that the sign is set by the spike-count structure of a recording at its temporal resolution, not by species. The thermodynamic signature of a neural code is therefore set not by the data alone but by the interaction order of the model used to read it.
\end{abstract}

\maketitle

The joint activity of $N$ neurons in a short time bin is a binary word $\mathbf{x}\in\{0,1\}^N$. The maximum-entropy distribution consistent with the measured firing rates and pairwise correlations is a pairwise Ising model~\cite{schneidman2006,shlens2006,tkacik2014}. Embedded in a family of distributions $P_T(\mathbf{x})\propto P(\mathbf{x})^{1/T}$ at fictitious temperature $T$, with $T=1$ the fitted model, this model has a specific-heat peak that approaches $T=1$ and grows with $N$~\cite{tkacik2015}. The pattern frequencies also follow Zipf's law~\cite{mora2011}. Both observations have been taken as evidence that neural populations operate near a critical point~\cite{tkacik2015}, and the reading has since been extended to other circuits~\cite{meshulam2019}.

\begin{figure*}[t]
\includegraphics[width=\textwidth]{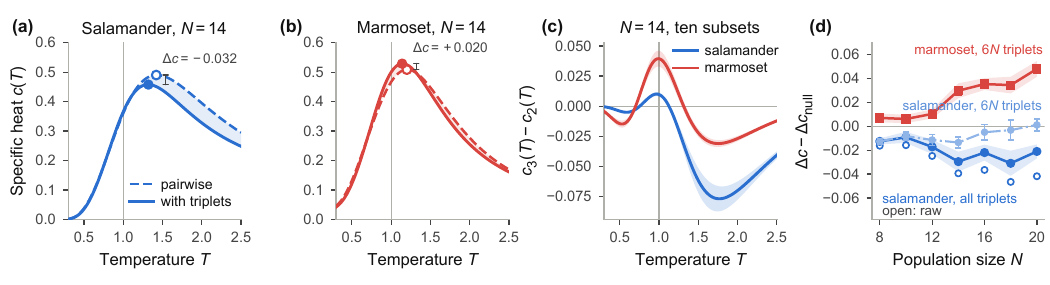}
\caption{Higher-order interactions renormalize the specific-heat peak with recording-dependent sign.
(a),(b) Fictitious-temperature specific heat per neuron, Eq.~\eqref{eq:cT}, of the exactly fitted pairwise (dashed) and third-order (solid) models of the $N=14$ subset whose $\Delta c$ is the median over the ten subsets, in salamander (a) and marmoset (b). The vertical line marks the operating point $T=1$.
(c) Difference between the third-order and pairwise specific heats, mean and s.e.m.\ over the ten $N=14$ subsets of each recording, with all triplets.
(d) Bias-corrected peak renormalization $\Delta c-\Delta c_{\rm null}$ versus population size for salamander with all triplets (blue circles) and $6N$ triplets (light blue), and for marmoset with $6N$ triplets (red squares). Mean and s.e.\ over ten subsets (thirteen for salamander $6N$ at $N=10$, eight for both $6N$ series at $N=18$ and $20$, six for salamander with all triplets at $N=20$). Open symbols: raw $\Delta c$, Eq.~\eqref{eq:dc}.}
\label{fig1}
\end{figure*}

That reading has been contested. Latent-variable models reproduce the signatures without fine-tuning~\cite{macke2011,schwab2014,aitchison2016,morrell2021}, subsampling a noncritical population produces a growing peak~\cite{nonnenmacher2017}, and inferred models sit generically close to critical points~\cite{mastromatteo2011}. Both sides work with the pairwise model, a truncation of the interaction expansion at second order, or with global extensions of it. Yet model studies show that higher-order structure changes the specific heat~\cite{macke2011,tkacik2015}. For about $100$ retinal ganglion cells responding to natural scenes, interactions beyond pairwise order are needed to describe the code~\cite{ganmor2011}, and higher-order correlations are documented in cortex~\cite{ohiorhenuan2010,yu2011}. Published thermodynamic analyses have used models of the number of active cells $K$ alone~\cite{tkacik2013}, the $K$-pairwise model, which constrains the distribution of $K$~\cite{tkacik2014,tkacik2015}, a global nonlinearity of the pairwise energy~\cite{humplik2017}, or homogeneous higher-order couplings shared by all cells~\cite{shimazaki2025}. To my knowledge, the thermodynamics of a model with explicit higher-order couplings fitted to the individual cells of a neural population has not been computed.

Does the thermodynamic signature of a neural code depend on the interaction order the code requires? Exact computation shows that it does. The third-order (triplet) terms that the data require change the peak by more than the finite-sample bias. The sign of the change differs between the two recordings studied, and the spike-count structure of the code sets it.

\textit{Models and thermodynamics.}
For a population of $N$ cells the models are maximum-entropy distributions of successive order,
\begin{equation}
P_k(\mathbf{x})=\frac{1}{Z_k}\exp\Big(\sum_{|S|\le k}\theta_S\prod_{i\in S}x_i\Big),
\label{eq:maxent}
\end{equation}
where $S$ runs over the nonempty subsets of $\{1,\dots,N\}$, $\theta_S$ are the couplings, and $Z_k$ is the partition function. The pairwise model is $k=2$ and the third-order model is $k=3$. The couplings with $|S|=3$ are the triplet terms. The $K$-pairwise model of Ref.~\cite{tkacik2014} is fitted as well. It adds to the pairwise model one parameter per value of the spike count $K=\sum_i x_i$, so that the distribution of $K$ is matched exactly. All models are fitted by maximum likelihood, with the model moments computed exactly by enumerating all $2^N$ states in blocks, which is practical up to $N=22$. An $L_2$ penalty of $10^{-6}$ per bin keeps the couplings of unobserved triplets finite. Two sets of triplet terms are used. The full third-order model includes all $\binom{N}{3}$ triplets and is practical up to $N=20$. The sparse set, of the kind used at large $N$~\cite{ganmor2011}, keeps the $6N$ triplets with the largest connected third-order correlation in the training data.

For each fitted model the energy is $E(\mathbf{x})=-\ln P_k(\mathbf{x})$, and the fictitious-temperature family $P_T\propto P_k^{1/T}$ has \mbox{specific} heat per neuron
\begin{equation}
c_k(T)=\frac{\mathrm{Var}_{P_T}[E]}{N\,T^2},
\label{eq:cT}
\end{equation}
evaluated exactly over all states on a grid of spacing $0.02$ in $T\in[0.3,2.5]$, with parabolic refinement of the maximum. The peak height $c_{k,\max}$ and its position $T^*_k$ are the quantities studied in Refs.~\cite{tkacik2015,mora2011}. The observable of this Letter is the shift of the peak produced by the triplet terms, its higher-order renormalization,
\begin{equation}
\Delta c = c_{3,\max}-c_{2,\max},
\label{eq:dc}
\end{equation}
computed for the same cells and the same training data, so that the comparison is not confounded by the choice or number of cells~\cite{nonnenmacher2017}. A surrogate introduced below tests whether the structure behind $\Delta c$ comes from cells outside the recorded population. The shift produced by the spike-count constraint alone, $\Delta c_K=c_{K,\max}-c_{2,\max}$, isolates the part set by the distribution of $K$.

Fitting a third-order model to finite data biases $\Delta c$ even when the data contain no higher-order structure, because the extra parameters absorb sampling noise. That bias is measured directly. At each $N$ a second, equally large set of cell subsets is drawn. For each subset, one to three synthetic data sets of the same length are sampled from its fitted pairwise model $P_2$. The entire procedure, including the selection of triplet terms, is repeated on these synthetic data, and their mean shift is $\Delta c_{\rm null}$. The synthetic data have no triplet interactions, so $\Delta c_{\rm null}$ is the finite-sample bias under that null. The correction assumes that the bias is the same at the true distribution. A check on data generated from fitted third-order models with known $\Delta c$ supports this: the corrected estimate recovers it to within $0.006$ (rms $0.003$ over six subsets), erring toward smaller salamander magnitudes. The bias-corrected renormalization $\Delta c-\Delta c_{\rm null}$ is a difference of independent means, quoted with the standard error (s.e.) of that difference. Throughout, $t$ is an estimate divided by its s.e., and a real subset counts as positive or negative when its shift lies above or below the mean bias. The bias of $\Delta c_K$ is below $0.003$ in every reported condition and below $0.001$ in both main natural-movie recordings.

\textit{Data and controls.}
The salamander data are the public recording of $160$ retinal ganglion cells responding to a natural movie~\cite{tkacik2014,marre2017}, binarized in $20$-ms bins into $2.8\times10^5$ population words with a mean firing probability of $0.036$ per bin. The marmoset data are the public recording of Refs.~\cite{sridhar2025,sridhar2026}, restricted to the $349$ ganglion cells with a mean firing probability above $0.01$ per bin. Their natural-movie responses are binarized in $23.5$-ms bins into $2.7\times10^5$ words with a mean firing probability of $0.091$ per bin. Training and held-out bins are separated by whole stimulus repeats (salamander) or by contiguous time blocks (marmoset). Each fit uses $1.5\times10^5$ random training bins, for ten random subsets of cells at each $N$. The triplet and $K$-pairwise models are scored against the pairwise model by their log-likelihood per bin on $5\times10^4$ held-out bins. Means and standard errors of the mean (s.e.m.) are over ten subsets unless stated otherwise.

\begin{figure*}[t]
\includegraphics[width=\textwidth]{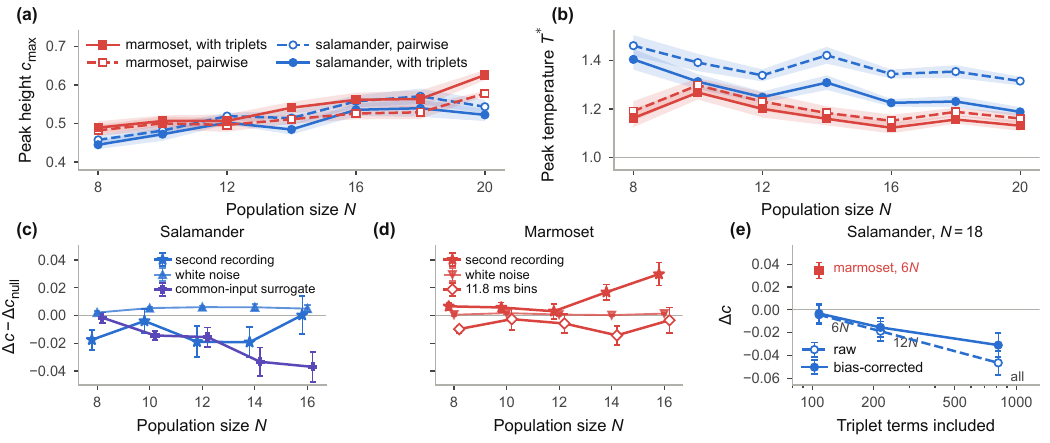}
\caption{Growth of the peak with $N$, controls, and dependence on the set of triplet terms.
(a) Peak height $c_{\max}$ versus $N$ for the pairwise (open, dashed) and bias-corrected third-order (filled, solid) models of salamander (blue, all triplets) and marmoset (red, $6N$ triplets), with s.e.m.\ bands.
(b) Peak temperature $T^*$ for the same models. The horizontal line marks the operating point $T=1$.
(c) Bias-corrected renormalization for the second salamander recording, salamander white noise ($6N$ triplets), and the common-input surrogate, offset horizontally for clarity.
(d) The same for the second marmoset recording, marmoset white noise ($6N$ triplets), and the marmoset data rebinned at $11.8$~ms.
(e) Salamander renormalization at $N=18$ against the number of triplet terms, raw (open) and bias-corrected (filled), with the bias-corrected marmoset $6N$ value at the same $N$.}
\label{fig2}
\end{figure*}

Four controls use the same fitting procedure, with the $6N$ triplets in the white-noise case. The first is a natural-movie recording from a second retina of each species. For salamander this is the $93$-cell population of Refs.~\cite{dryad2024,hoshal2024} in $16.7$-ms bins ($54$ cells with mean firing probability above $0.005$ per bin, mean $0.016$). For marmoset it is a second retina of Ref.~\cite{sridhar2025} in $23.5$-ms bins ($163$ cells, mean $0.064$, $9.6\times10^4$ training bins). The second is white noise, under which the higher-order structure of the \mbox{salamander} code is weak~\cite{ganmor2011}: the $93$-cell \mbox{salamander} population in $33$-ms bins ($8.2\times10^4$ training bins, mean firing probability $0.067$ per bin) and the first marmoset recording ($5.0\times10^4$ bins). The third separates temporal resolution from species: the marmoset data rebinned at their native $11.8$-ms frame, which lowers the mean firing probability per bin to $0.056$, nearer the salamander $0.036$. The fourth is a common-input surrogate for each salamander subset: a dichotomized Gaussian~\cite{macke2009} with one shared input, matched to the measured rates and mean pairwise correlation ($10^5$ training and $3\times10^4$ held-out bins). Such models reproduce apparent criticality without explicit couplings~\cite{macke2011,schwab2014}.

\textit{Results.}
Figures~\ref{fig1}(a) and \ref{fig1}(b) show the effect for one subset from each recording. In salamander the triplet terms lower the peak, and in marmoset they raise it. The difference $c_3(T)-c_2(T)$ has the same shape in both recordings [Fig.~\ref{fig1}(c)]. It is positive in a window around the operating point $T=1$, from $T\simeq0.7$ up to a crossing at $T\simeq1.1$ to $1.3$, and negative above the crossing. In both recordings the triplet terms thus raise the specific heat at the operating point and pull the peak toward it. Since $c_2(T^*_2)\ge c_2(T^*_3)$ and $c_3(T^*_3)\ge c_3(T^*_2)$, the shift is bounded by $c_3(T^*_2)-c_2(T^*_2)\le\Delta c\le c_3(T^*_3)-c_2(T^*_3)$. It is negative whenever the third-order peak lies above the crossing, as both peaks do in salamander ($T^*\simeq1.3$ to $1.4$ at $N=14$), and positive whenever the pairwise peak lies below it, as both do in marmoset ($1.1$ to $1.2$). With all triplets the raw salamander shift is negative at every $N$ and grows in magnitude with $N$ [Fig.~\ref{fig1}(d)], and so does the bias $\Delta c_{\rm null}$, which tracks the number of triplet parameters. The bias-corrected shift (Table~\ref{tab1}) stays negative at every $N$, with $t$ between $-2.0$ and $-5.0$. Pooled over all $66$ subsets it is $-0.020$ with $t=-7.8$ ($60$ of $66$ subsets negative). In marmoset the bias-corrected shift is small for $N\le12$ and then grows, reaching $+0.048$ at $N=20$ with the $6N$ triplets. The pooled value is $+0.024$ with $t=12$ ($59$ of $66$ positive). The triplet terms are needed in both recordings, since their held-out log-likelihood gain is positive at every $N$ and largest at $N=20$ \mbox{(Table~\ref{tab1})}.

\begin{table}[b]
\caption{Bias-corrected peak renormalization by the triplet terms, $\Delta c-\Delta c_{\rm null}$, with its s.e.\ in parentheses in units of the last digit, the shift produced by the spike-count constraint alone, $\Delta c_K$, and the held-out log-likelihood gain of the third-order model over the pairwise model, $\Delta\ell$, in nats per bin. Entries are means over ten subsets with all triplets, except that the marmoset entries in the rows marked $\dagger$ use $6N$ triplets and eight subsets, and the salamander entries in the row marked $\ddagger$ use six subsets.}
\label{tab1}
\begin{ruledtabular}
\begin{tabular}{cccc@{\hspace{1.2em}}ccc}
 & \multicolumn{3}{c}{Salamander} & \multicolumn{3}{c}{Marmoset}\\
$N$ & $\Delta c-\Delta c_{\rm null}$ & $\Delta c_K$ & $\Delta\ell$ & $\Delta c-\Delta c_{\rm null}$ & $\Delta c_K$ & $\Delta\ell$\\
\hline
8  & $-0.013(3)$ & $-0.011$ & $0.002$ & $+0.007(4)$ & $+0.004$ & $0.003$\\
10 & $-0.009(5)$ & $-0.011$ & $0.006$ & $+0.000(5)$ & $-0.005$ & $0.004$\\
12 & $-0.017(6)$ & $-0.015$ & $0.011$ & $+0.007(6)$ & $-0.003$ & $0.009$\\
14 & $-0.030(8)$ & $-0.028$ & $0.010$ & $+0.031(9)$ & $+0.016$ & $0.018$\\
16 & $-0.022(7)$ & $-0.026$ & $0.019$ & $+0.040(8)$ & $+0.024$ & $0.027$\\
18 & $-0.031(11)$ & $-0.028$ & $0.027$ & $+0.035(7)^{\dagger}$ & $+0.020$ & $0.021$\\
20 & $-0.021(6)^{\ddagger}$ & $-0.026$ & $0.031$ & $+0.048(7)^{\dagger}$ & $+0.036$ & $0.029$\\
\end{tabular}
\end{ruledtabular}
\end{table}

Figure~\ref{fig2}(a) shows the growth of the peak with $N$, the quantity at the center of the debate. In salamander the bias-corrected third-order peak grows more slowly than the pairwise peak: the ratio of their slopes $dc_{\max}/dN$ is $0.84$, with a $95\%$ bootstrap interval over subsets of $0.71$ to $0.96$. In marmoset it grows faster, with a ratio of $1.55$ ($95\%$ interval $1.39$ to $1.83$) for the $6N$ triplets over $N=8$ to $20$ and $1.98$ ($1.4$ to $5.5$) for all triplets over $N=8$ to $16$. Figure~\ref{fig2}(b) shows the peak temperature. In salamander the triplet terms move it toward the operating point, with $T^*_2\simeq1.3$ to $1.5$ and $T^*_3\simeq1.2$ to $1.4$, while in marmoset both models give $T^*\simeq1.1$ to $1.3$. For the spike-count constraint, Ref.~\cite{tkacik2015} gives the pairwise and $K$-pairwise peaks of the salamander recording at $N=20$ to $120$ (its Fig.~S2). The $K$-pairwise peak is lower at $N=20$, as here, but higher from $N=40$ on. The sign of $\Delta c_K$ thus depends on $N$ as well, as expected when the pairwise errors in $P(K)$ change with $N$~\cite{tkacik2014}.

\begin{figure}[t]
\includegraphics[width=\columnwidth]{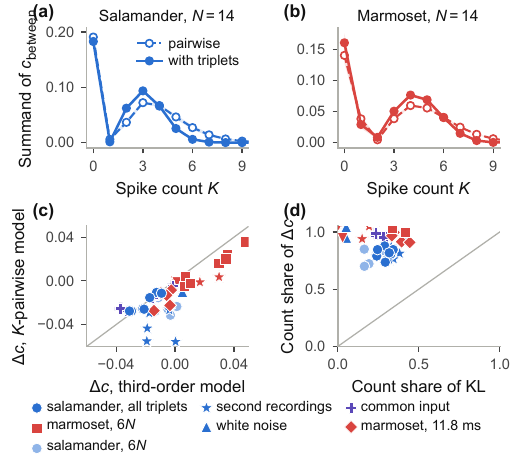}
\caption{The renormalization is a change in the energy variance across spike counts.
(a),(b) Summand $P_T(K)[\langle E\,|\,K\rangle-\langle E\rangle]^2/NT^2$ of $c_{\rm between}$, Eq.~\eqref{eq:dec}, versus spike count $K$ for the pairwise (open) and third-order (filled) models at $N=14$, mean over ten subsets with s.e.m.\ smaller than the symbols, each model at its own $T^*$.
(c) Shift from the spike-count constraint alone, $\Delta c_K$, against the bias-corrected renormalization from the triplet terms, for every condition and $N$, with the identity line.
(d) Fraction of $\Delta c$ in the count term against the fraction of the Kullback--Leibler divergence between the models that lies in the count distribution, with the identity line. Blue, salamander. Red, marmoset. Violet, surrogate. Symbols as in the legend below.}
\label{fig3}
\end{figure}

The controls in Figs.~\ref{fig2}(c) and \ref{fig2}(d) locate the effect. The second marmoset recording reproduces the positive shift at every $N$ (pooled $t=5.1$). The second salamander recording is far sparser, and there the full third-order model is bias-limited: at $N=16$ the bias equals the raw shift. The spike-count constraint, whose bias is below $0.003$, is the reliable probe there. It gives the salamander sign at every $N$, with $\Delta c_K=-0.027$ to $-0.056$, about twice the value in the first recording. The $6N$ triplets give a large positive value (see below). Under white noise the bias-corrected shift with the $6N$ triplets is at most $+0.006$ in either recording, against $-0.013$ to $-0.005$ for the salamander natural movie over the same $N$. There the held-out gain of the triplet terms is at most $0.001$ nats per bin. Rebinning the marmoset data at $11.8$~ms reverses the sign, for the third-order model (pooled $t=-2.2$) and for the spike-count constraint alike [Fig.~\ref{fig3}(c)]. The common-input surrogate reproduces the salamander sign (pooled $t=-5.9$), even though the triplet terms do not improve its held-out likelihood at all. In every natural-movie and surrogate condition, $c_3(T)-c_2(T)$ at $N=14$ has the shape of Fig.~\ref{fig1}(c): positive at $T=1$ by at least three standard errors, and crossing zero between $T=1.1$ and $1.4$. Both mean peaks lie above the crossing in the four negative cases and below it in the two positive cases, as the bound requires.

\textit{Dependence on the set of triplet terms.}
In marmoset the $6N$ triplets reproduce the full third-order shift within the combined s.e.m.\ at every $N$. In salamander they do not. The $6N$ shift falls from $100\%$ of the full third-order shift at $N=8$ to $22\%$ at $N=16$ and is consistent with zero beyond. At $N=18$ the bias-corrected salamander shift decreases monotonically with the number of triplet terms: $-0.003\pm0.008$ with the $6N$ triplets, $-0.015\pm0.008$ with the $12N$ triplets selected by the same criterion, and $-0.031\pm0.011$ with all $816$ triplets [Fig.~\ref{fig2}(e)]. In the second salamander recording the $6N$ triplets even reverse the sign. The salamander effect is spread over many weak terms that a sparse selection misses, whereas in marmoset the selected terms carry it.

\textit{Mechanism.}
The specific heat has an exact decomposition by population spike count. By the law of total variance,
\begin{equation}
c(T)=\underbrace{\frac{\langle \mathrm{Var}[E\,|\,K]\rangle_K}{N T^2}}_{c_{\rm within}}+\underbrace{\frac{\mathrm{Var}_K[\langle E\,|\,K\rangle]}{N T^2}}_{c_{\rm between}},
\label{eq:dec}
\end{equation}
where the averages are over $P_T$. The first term is the spread of energy among patterns with the same number of active cells (which cells fire together), the second the variance of energy across spike counts (how many fire together). Evaluated at each model's own $T^*$, the decomposition splits $\Delta c$ into $\Delta c_{\rm within}+\Delta c_{\rm between}$ to within the $10^{-4}$ accuracy of the peak refinement.

The count term $c_{\rm between}$ dominates the shift in every condition [Fig.~\ref{fig3}(d)]. It accounts for $74$ to $88\%$ of the raw $\Delta c$ in salamander and $85$ to $116\%$ in marmoset (subset medians). In salamander the triplet terms lower the contribution of the high-count tail more than they raise that of $K=2$ to $3$ [Fig.~\ref{fig3}(a)]. In marmoset they raise the contributions of the silent state and of moderate counts [Fig.~\ref{fig3}(b)]. Constraining the distribution of $K$ alone, with no triplet terms, reproduces the whole bias-corrected salamander shift ($84$ to $123\%$ at every $N$) and $53$ to $74\%$ of the marmoset shift for $N\ge14$ [Fig.~\ref{fig3}(c)].

The decomposition also explains why the thermodynamic effect is not proportional to the likelihood gain. Only $22$ to $43\%$ of the Kullback--Leibler divergence between the two models lies in the count distribution [Fig.~\ref{fig3}(d)]. Most of the probability that the triplet terms move is redistributed among patterns of equal count, which barely changes the specific heat. The likelihood responds mainly to which cells fire together, the specific heat mainly to how many.

\textit{Discussion.}
Three conclusions follow. First, the thermodynamic signature read as criticality in pairwise models depends on the interaction order the code requires. The triplet terms change the peak height and its growth with $N$ by more than the finite-sample bias, and within the $6N$ set the change nearly vanishes under white noise. Second, the change is dominated by the count term of Eq.~\eqref{eq:dec}, and its sign is set by whether the required higher-order terms reduce or increase the variance of the energy across spike counts. In salamander the pairwise model predicts synchronous events that the triplet terms remove. By the peak-height criterion of Ref.~\cite{tkacik2015}, the apparent criticality of the pairwise model is then overstated at these population sizes. By peak position, and at $T=1$ itself, the third-order model is nonetheless closer to criticality. In marmoset, by contrast, the apparent criticality is understated. The same marmoset cells rebinned at $11.8$~ms show the salamander sign, and a common-input surrogate reproduces it. The sign thus belongs to the spike-count structure of a recording at its temporal resolution, not to a species. As the surrogate shows, this sign needs no explicit higher-order couplings, only a spike-count structure that the pairwise model cannot represent. Third, the spike-count constraint of the $K$-pairwise model~\cite{tkacik2014} captures the whole correction in salamander, so at these population sizes the $K$-pairwise peak of Ref.~\cite{tkacik2015} already includes it. In marmoset it captures only half to three quarters of the correction. Fourth-order terms were not fitted. Since the spike-count constraint contains cell-symmetric terms of every order, any effect beyond third order that it misses must be cell-specific, and the marmoset is where to look for it.

The pairwise-model thermodynamics of a neural population therefore cannot be interpreted without the higher-order terms the population requires. Beyond the $N\simeq22$ reach of enumeration, sampling-based estimators of the partition function will be needed at the hundreds of cells where higher-order structure dominates the code.

\begin{acknowledgments}
The data that support the findings of this article are openly available in Refs.~\cite{marre2017,dryad2024,sridhar2025}.
\end{acknowledgments}

\sloppy


\begin{thebibliography}{99}
\bibitem{schneidman2006} E.~Schneidman, M.~J.~Berry II, R.~Segev, and W.~Bialek, Nature \textbf{440}, 1007 (2006).
\bibitem{shlens2006} J.~Shlens, G.~D.~Field, J.~L.~Gauthier, M.~I.~Grivich, D.~Petrusca, A.~Sher, A.~M.~Litke, and E.~J.~Chichilnisky, J. Neurosci. \textbf{26}, 8254 (2006).
\bibitem{tkacik2014} G.~Tka\v{c}ik, O.~Marre, D.~Amodei, E.~Schneidman, W.~Bialek, and M.~J.~Berry II, PLoS Comput. Biol. \textbf{10}, e1003408 (2014).
\bibitem{tkacik2015} G.~Tka\v{c}ik, T.~Mora, O.~Marre, D.~Amodei, S.~E.~Palmer, M.~J.~Berry II, and W.~Bialek, Proc. Natl. Acad. Sci. U.S.A. \textbf{112}, 11508 (2015).
\bibitem{mora2011} T.~Mora and W.~Bialek, J. Stat. Phys. \textbf{144}, 268 (2011).
\bibitem{meshulam2019} L.~Meshulam, J.~L.~Gauthier, C.~D.~Brody, D.~W.~Tank, and W.~Bialek, Phys. Rev. Lett. \textbf{123}, 178103 (2019).
\bibitem{macke2011} J.~H.~Macke, M.~Opper, and M.~Bethge, Phys. Rev. Lett. \textbf{106}, 208102 (2011).
\bibitem{schwab2014} D.~J.~Schwab, I.~Nemenman, and P.~Mehta, Phys. Rev. Lett. \textbf{113}, 068102 (2014).
\bibitem{aitchison2016} L.~Aitchison, N.~Corradi, and P.~E.~Latham, PLoS Comput. Biol. \textbf{12}, e1005110 (2016).
\bibitem{morrell2021} M.~C.~Morrell, A.~J.~Sederberg, and I.~Nemenman, Phys. Rev. Lett. \textbf{126}, 118302 (2021).
\bibitem{nonnenmacher2017} M.~Nonnenmacher, C.~Behrens, P.~Berens, M.~Bethge, and J.~H.~Macke, PLoS Comput. Biol. \textbf{13}, e1005718 (2017).
\bibitem{mastromatteo2011} I.~Mastromatteo and M.~Marsili, J. Stat. Mech. P10012 (2011).
\bibitem{ganmor2011} E.~Ganmor, R.~Segev, and E.~Schneidman, Proc. Natl. Acad. Sci. U.S.A. \textbf{108}, 9679 (2011).
\bibitem{ohiorhenuan2010} I.~E.~Ohiorhenuan, F.~Mechler, K.~P.~Purpura, A.~M.~Schmid, Q.~Hu, and J.~D.~Victor, Nature \textbf{466}, 617 (2010).
\bibitem{yu2011} S.~Yu, H.~Yang, H.~Nakahara, G.~S.~Santos, D.~Nikoli\'c, and D.~Plenz, J. Neurosci. \textbf{31}, 17514 (2011).
\bibitem{tkacik2013} G.~Tka\v{c}ik, O.~Marre, T.~Mora, D.~Amodei, M.~J.~Berry II, and W.~Bialek, J. Stat. Mech. P03011 (2013).
\bibitem{humplik2017} J.~Humplik and G.~Tka\v{c}ik, PLoS Comput. Biol. \textbf{13}, e1005763 (2017).
\bibitem{shimazaki2025} U.~Rodr\'iguez-Dom\'inguez and H.~Shimazaki, Neural Comput. \textbf{37}, 2011 (2025).
\bibitem{marre2017} O.~Marre, G.~Tka\v{c}ik, D.~Amodei, E.~Schneidman, W.~Bialek, and M.~J.~Berry II, Multi-electrode array recording from salamander retinal ganglion cells, IST Austria (2017), \href{https://doi.org/10.15479/AT:ISTA:61}{doi:10.15479/AT:ISTA:61}.
\bibitem{sridhar2025} S.~Sridhar and T.~Gollisch, Marmoset retinal ganglion cell responses to naturalistic movies and spatiotemporal white noise, G-Node (2025), \href{https://doi.org/10.12751/g-node.t43ph1}{doi:10.12751/g-node.t43ph1}.
\bibitem{sridhar2026} S.~Sridhar, M.~Vystr\v{c}ilov\'a, M.~H.~Khani, D.~Karamanlis, H.~M.~Schreyer, V.~Ramakrishna, S.~Kr\"uppel, S.~J.~Zapp, M.~Mietsch, A.~S.~Ecker, and T.~Gollisch, PLoS Comput. Biol. \textbf{22}, e1014157 (2026).
\bibitem{dryad2024} S.~E.~Palmer, C.~Holmes, B.~Hoshal, K.~Bojanek, J.~Salisbury, M.~J.~Berry II, and O.~Marre, Larval salamander retinal population data in response to natural movies from the Chicago Motion Database, Dryad (2024), \href{https://doi.org/10.5061/dryad.4qrfj6qm8}{doi:10.5061/dryad.4qrfj6qm8}.
\bibitem{hoshal2024} B.~D.~Hoshal, C.~M.~Holmes, K.~Bojanek, J.~M.~Salisbury, M.~J.~Berry II, O.~Marre, and S.~E.~Palmer, Proc. Natl. Acad. Sci. U.S.A. \textbf{121}, e2313676121 (2024).
\bibitem{macke2009} J.~H.~Macke, P.~Berens, A.~S.~Ecker, A.~S.~Tolias, and M.~Bethge, Neural Comput. \textbf{21}, 397 (2009).
\end{thebibliography}
\end{document}